# Quantum-Confined Electronic States arising from Moiré Pattern of MoS$_2$-WSe$_2$ Hetero-bilayers

Yi Pan,[1,2] Stefan Fölsch,[1] Yifan Nie,[3] Dacen Waters,[4] Yu-Chuan Lin,[5] Bhakti Jariwala,[5] Kehao Zhang,[5] Kyeongjae Cho,[3] Joshua A. Robinson,[5] and Randall M. Feenstra[4]

[1]Paul-Drude-Institut für Festkörperelektronik, Hausvogteiplatz 5-7, 10117 Berlin, Germany
[2]Center for Spintronics and Quantum Systems, State Key Laboratory for Mechanical Behavior of Materials, Xi'an Jiaotong University, Xi'an 710049, China
[3]Dept. Materials Science and Engineering, The University of Texas at Dallas, Dallas, Texas, 75080 U.S.A.
[4]Dept. Physics, Carnegie Mellon University, Pittsburgh, PA, 15213 U.S.A.
[5]Dept. Materials Science and Engineering, and Center for 2-Dimensional and Layered Materials, The Pennsylvania State University, University Park, PA, 16802 U.S.A.

**ABSTRACT:** A two-dimensional (2D) hetero-bilayer system consisting of MoS$_2$ on WSe$_2$, deposited on epitaxial graphene, is studied by scanning tunneling microscopy and spectroscopy at temperatures of 5 and 80 K. A moiré pattern is observed, arising from lattice mismatch of 3.7% between the MoS$_2$ and WSe$_2$. Significant energy shifts are observed in tunneling spectra observed at the maxima of the moiré corrugation, as compared with spectra obtained at corrugation minima, consistent with prior work. Furthermore, at the minima of the moiré corrugation, sharp peaks in the spectra at energies near the band edges are observed, for spectra acquired at 5 K. The peaks correspond to discrete states that are confined within the moiré unit cells. Conductance mapping is employed to reveal the detailed structure of the wave functions of the states. For measurements at 80 K, the sharp peaks in the spectra are absent, and conductance maps of the band edges reveal little structure.

Vertical heterostructures of various two-dimensional (2D) materials have been studied intensely over the past decade due to their novel electronic and optical properties.[1,2,3,4] Just as different electronic properties are found between 1 monolayer (ML) and 2 ML of a *given* material (e.g. MoS$_2$),[2,5] due to hybridization of the electronic states between the layers, so too can a combination of two *different* materials (such as MoS$_2$ on WSe$_2$) be expected to produce electronic states that are not a simple combination of the states of the constituent materials.[6,7,8,9,10,11,12] Understanding in detail the hybridization effects that occur between specific 2D layers constitutes a very important topic, such that we might obtain some general understanding (and predictive capability) for arbitrary 2D layers that are combined together into a heterostructure.[13,14,15]

When two MLs of different 2D materials are combined to form a hetero-bilayer (or when rotational misalignment occurs between the lattice of MLs of the same 2D material), then a moiré pattern will form.[16] Such patterns have been studied in detail e.g. for graphene on h-BN,[17,18,19,20] and also for various transition-metal dichalcogenide (TMD) materials.[7,18,21,22] In a recent report, Zhang et al. described scanning tunneling microscopy and spectroscopy (STM/STS) obtained for hetero-bilayers of MoS$_2$ and WSe$_2$, with the measurements performed at a temperature near 77 K.[7] A moiré corrugation with relatively large amplitude of 0.17 nm (sample bias +3 V) and period of 8.7±0.2 nm was observed. Significantly, shifts in the band-edge locations as large as 0.2 eV were found between tunneling spectra obtained at the maxima of this corrugation compared to spectra obtained from the two different kinds of minima. It was argued that hybridization of WSe$_2$-derived valence band (VB) states at the Γ-point in **k**-space, with resulting energy being quite sensitive to the MoS$_2$-WSe$_2$ separation, was a large contributor to the observed variation in the band-edge energies.

In our work we also study vertical hetero-bilayers of MoS$_2$ on WSe$_2$, grown in our case by a combination of powder-vaporization chemical vapor deposition (CVD, for MoS$_2$) and metal-organic chemical vapor deposition (MOCVD, for WSe$_2$).[8,23,24,25] STM/STS measurements were performed in a low-temperature STM system operated at a base temperature of 5 K. We employ a variable-$z$ measurement method for the STS which ensures high dynamic range in the conductance (see Supplemental Information for details).[26] The STM/STS results that we obtain are very similar in many respects to those of Zhang et al.[7] In particular, we find a moiré corrugation with amplitude of 0.13 nm (sample bias +1.5 V) and period of 8.5±0.2 nm, and spectral shifts as large as 0.2 eV are observed between tunneling spectra obtained at maxima of the corrugation compared with those obtained at the two inequivalent types of minima.

Additionally, for measurements performed at 5 K, we also observe narrow (< 10 meV), sharp peaks that occur near band-edge energies of the spectra, for both the valence band (VB) and the conduction band (CB). We argue that such states arise from quantum confinement in the spatially varying potential associated with the moiré pattern. Significant variation in the peak position occurs between different moiré cells, likely arising from a randomly varying potential arising from point defects in the material. At the higher temperature of 80 K, we find that the sharp peaks associated with the localized states disappear, although the energy shifts between spectra obtained at corrugation minima and maxima remain.

Figure 1 shows STM images of the MoS$_2$-WSe$_2$ hetero-bilayer, showing its moiré pattern with maxima in the corrugation and two types of minima. We label the spatial locations corresponding to corrugation maxima as A, the minima that are deepest for sample bias of −1.5 V but intermediate in depth for at +1.5 V as B, and the other type of minima as C (see Figs. 1(b,c,d)). The appearance of the moiré corrugation is completely consistent with the prior work of Zhang et al.[7] Those workers demonstrated on the basis of annular dark-field scanning transmission electron microscopy that their CVD-grown material had the MoS$_2$ and WSe$_2$ layers stacked with 0° rotational misalignment (as opposed to 180° or some other angle).[7] For our samples, we find the majority of the sample to have rotational misalignment of either 0° (R-stacking) or 180° (H-stacking), as described in the Supplemental Material, Fig. S1. The 0° and 180° cases cannot be distinguished on the basis of STM images alone,[7] and hence we tentatively utilize the 0° angle determined by Zhang et al. We observe a moiré period of 8.5±0.2 nm, which is consistent with lattice constants of 0.316 and 0.328 nm for $MoS_2$ and $WSe_2$, respectively, so that 27 unit cells of $MoS_2$ fit onto 26 unit cells of $WSe_2$. Comparing both the STS data presented below and voltage-dependent images shown in Fig. S2 to the data of Zhang et al.,[7] the locations that we label A, B, and C are found to correspond, in their notation, to AA, $AB_W$, and $AB_{Se}$, respectively. At the A=AA locations, the Mo atoms are directly over the W and the S atoms are directly over the Se; at the B=$AB_W$ locations, the Mo atoms are over the Se atoms and the W atoms are visible through the $MoS_2$; at the C=$AB_{Se}$ locations, the S atoms are over the W atoms and the Se atoms are visible through the $MoS_2$ (as shown in Fig. 1(e), and with a full view of moiré unit cell shown in Fig. S3). First-principles computations reveal a 0.06 nm difference in the equilibrium separation of $MoS_2$ and $WSe_2$ for these various registries, as listed in Table 1 (and in agreement with Zhang et al.[7]). Associated with the differing registries, the theoretical energies of band edges are found to change, as also shown in Table 1 (computed using the Vienna Ab-Initio Simulation Package[27] with the projector-augmented wave method,[28] employing the Purdew-Burke-Ernzerhof generalized gradient approximation exchange-correlation functional[29] together with dipole corrections obtained by Grimme's DFT-D2 method,[30] as further detailed in the Supplementary Information). Notation for the labeling of bands is similar to that employed by Zhang et al., according to the point in **k**-space that the band is centered on (Γ, K, or Q) and the layer (W for $WSe_2$ or M for $MoS_2$) from which the band originates.[5,7]

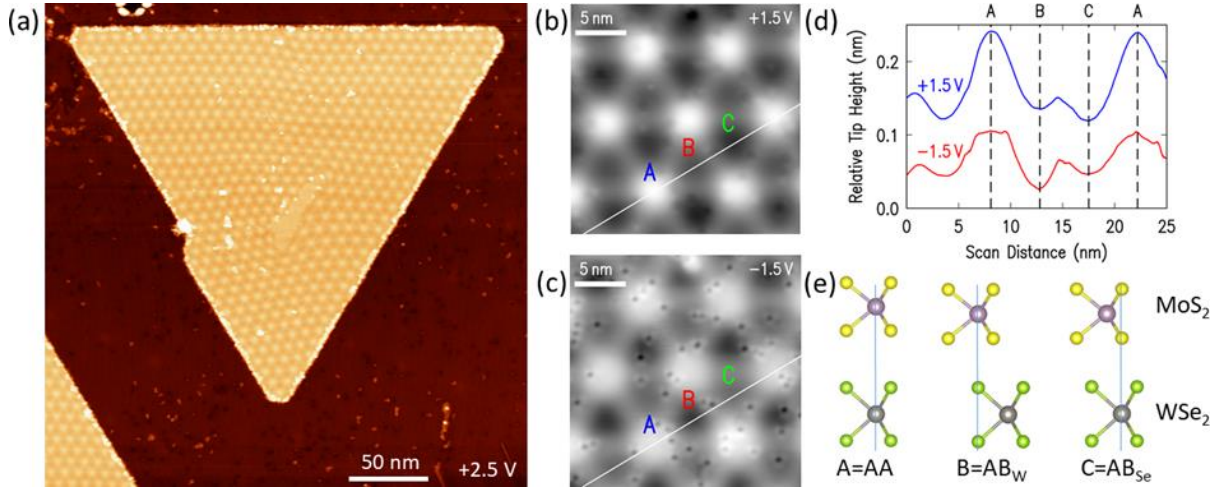

FIG 1. STM data acquired at 5 K, showing (a) large-area image, (b) and (c) images acquired at +1.5 and -1.5 V, (d) cross-sectional cuts of inset images, (e) schematic view of respective registrations between MoS$_2$ and WSe$_2$; cases labeled A, B, and C correspond, using structural analysis and notation from Ref. [7], to AA, AB$_W$ and AB$_{Se}$.

| registry | separation (nm) | E (eV) – $E_{VAC,W}$ | |
|---|---|---|---|
| | | $Γ_W$ | $K_M$ |
| A=AA | 0.690 | -5.29 | -4.57 |
| B=$AB_W$ | 0.632 | -5.09 | -4.55 |
| C=$AB_{Se}$ | 0.629 | -5.04 | -4.46 |

Table I. First-principles computational results for various registries (A, B, C) of 1×1 unit cell of $MoS_2$ on $WSe_2$, listing the equilibrium $MoS_2$-$WSe_2$ separation and the energies of the $Γ_W$ VB and the $K_M$ CB edges relative to the vacuum level on the $WSe_2$ side of the $MoS_2$-$WSe_2$ bilayer.

Figure 2(b) shows typical spectra obtained from the $MoS_2$-$WSe_2$ hetero-bilayer, which can be compared to spectra obtained from individual, isolated layers of $MoS_2$ on epitaxial graphene (EG) and $WSe_2$ on EG as shown in Fig. S4. The highest lying valence band (VB) of the hetero-bilayer is labeled $Γ_W$; this band derives primarily from the $WSe_2$. It is important to note that a higher VB also exists, centered at the K point and also associated with $WSe_2$ ($K_W$ band). We can observe that band in spectra of an isolated $WSe_2$ layer on EG and also for signal-averaged spectra of the hetero-bilayer (Fig. S4) but it is not visible (due to low intensity) in Fig. 2 since it originates both from an edge point of the Brillouin zone[31,32] and from the $WSe_2$ layer that is *beneath* the $MoS_2$. The lowest lying conduction band (CB) in Fig. 2(a) is labeled $K_M$, deriving primarily from a $MoS_2$ band centered at the K point of the BZ. Significant differences are seen in the spectra of Fig. 2(b) depending on the location within the moiré unit cell that they are acquired at. For spectra acquired from the A locations (corrugation maxima), we find results similar to those previously presented by Zhang et al.[7] However, if we look to other locations in the unit cell, we obtain significantly different results. Specifically, examining spectra from locations B and C in Fig. 2(b), we find intense, sharp peaks observed at energies near the band edges of the $\Gamma_W$ VB and $K_M$ CB. This type of sharp spectral feature from a TMD hetero-bilayer has not been previously reported.

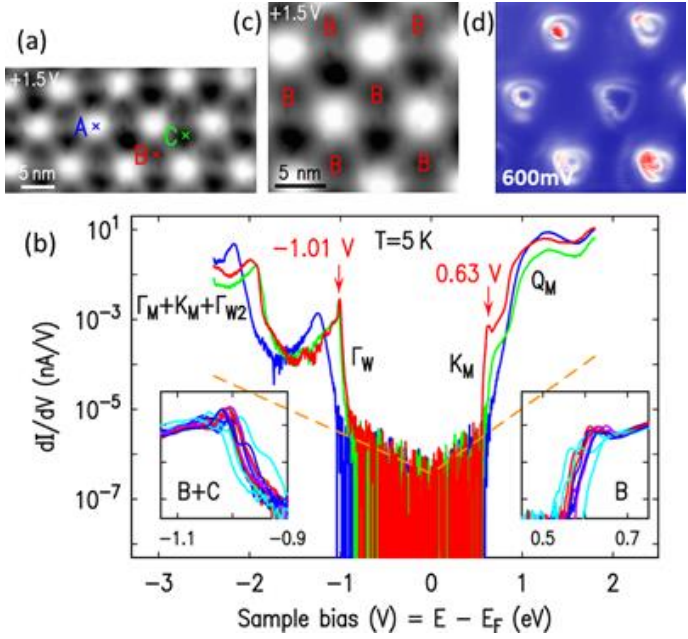

FIG 2. STM/STS data acquired at 5 K showing (a) constant-current image of hetero-bilayer of $MoS_2$ on $WSe_2$, (b) selected spectra acquired at the points indicated in (a), with insets showing expanded view of band-edge peaks and including data from nearby spatial locations, (c) constant-current image of hetero-bilayer, and (d) constant-height conductance map acquired at 600 mV (same color range as in Fig. 3(b)).

Expanded views of the sharp band-edge peaks are provided in the insets of Fig. 2(b), showing spectra obtained from B and C locations (negative bias) or from C locations only (positive bias); additional band-edge spectra are displayed in Fig. S5. At the B locations we find sharp peaks at both the $\Gamma_W$ VB and $K_M$ CB edges, whereas at the C locations we only find sharp peaks at the $\Gamma_W$ VB edge. We use a modulation voltage of $V_{mod} = 10$ mV rms in the measurements. The corresponding energy resolution[33] (full-width at half-maximum, fwhm) is given by $\Delta E = \sqrt{(3.5kT)^2 + (2.5eV_{mod})^2} = 25.0$ meV at $T = 5$ K. Many of the $\Gamma_W$ VB peaks (left inset) are seen to have width very close to this value. Hence, these peaks have intrinsic width considerably less than 10 mV, with the modulation producing the observed width. The dominant peak for these $\Gamma_W$ VB features lies typically at −1.01 V, with a spread of 10 – 20 mV in the position of this peak, comparing spectra from different moiré cells. (Our measurement reproducibility for peak positions measured repeatedly at the *same* location is <1 meV). In contrast, for the $K_M$ CB peaks (right inset) the width of some of these might be limited by the 10 mV modulation, but others appear to be broader. Additionally, there are multiple band-edge peaks in many of those spectra. Nevertheless, closely studying these spectra we again can see cell-to-cell variation of 10 – 20 mV in the position of specific peaks in the spectra. We attribute the variation in peak position for both the $\Gamma_W$ and $K_M$ peaks to the influence of a randomly varying potential in the layers, likely due to the point defects in the material (as seen e.g. in Fig. 1(c), and discussed in Ref. [25]; some defects exhibit midgap states and others do not, but all spectra presented in the present work are acquired sufficiently far from any defect such that no defect states appear in the results). With the width of the spectral peaks being less than their cell-to-cell variation in position, we speculate that it is quite possible that the corresponding states are spatially localized within individual moiré cells (transport through the states would likely still be enabled by their connection to the graphene below the hetero-bilayer).

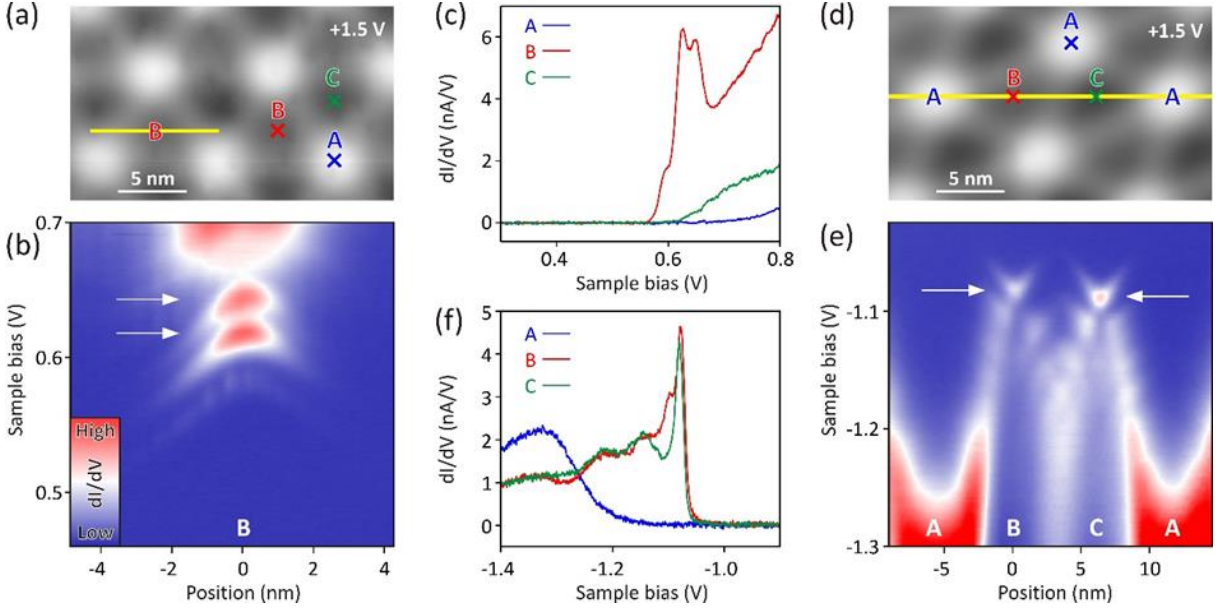

FIG 3. (a) STM image, with moiré locations A, B, and C indicated. (b) Constant-height conductance map taken along the yellow line in (a) for voltages in the conduction band-edge region, revealing two B-confined states (marked by arrows) and the band onset at higher energy (broader B-centered conductance feature). (c) Tunneling spectra recorded as indicated by crosses in (a). (d) Same as (a) but at a different area of the $MoS_2$-$WSe_2$ moiré structure. (e) Constant-height conductance map taken along the yellow line in (d) for voltages in the valence band-edge region; confined states occur at locations B and C. (f) Tunneling spectra recorded as indicated by crosses in (d). Spectra in (c) and (f) are recorded with the variable-$z$ method, but are not normalized to constant $z$.

Associated with these sharp features we observe electronic states that are spatially confined within the moiré unit cells, an example of which is shown in Figs. 2(c) and 2(d). Figure 2(c) shows a constant-current topography image with the B locations marked, and Fig. 2(d) shows a constant-height conductance map of the same surface area, acquired at +0.6 V. Distinct rings are seen, confined in the area of the type-B corrugation minima; we associate these features (and the sharp band-edge peaks) with quantum-confined states in the moiré unit cell. At higher voltages, these rings evolve into more extended features within the moiré pattern, associated with extended band states, similar to those presented by Zhang et al.[7] Figure S5 presents additional conductance maps, over a wide range of bias voltages.

Additional information on the spatial arrangement of the quantum-confined states near the band edges is obtained by conductance mapping as a function of both energy and spatial position, as shown in Fig. 3 for both the $\Gamma_W$ VB and $K_M$ CB band-edge states. Starting our discussion with the CB states, the conductance was probed with the tip held at constant height and at a given bias along a line spanning a location B as indicated by the yellow line in Fig. 3(a). Performing this measurement as a function of sample bias yields the spatial conductance map in Fig. 3(b). Two

B-confined states are found which are separated by ~30 meV. The V-shaped streaks branching off from the conductance maxima are likely due to tip-induced band bending (TIBB) pushing up the confined-state energy when the tip approaches the location B. The TIBB effect explains also the ring features shown in Fig. 2(d) and their evolution with bias voltage (Fig. S6). The spectra in Fig. 3(c) are consistent with the conductance map and reveal a double peak plus a shoulder when the tip probes location B, whereas no confined states occur at locations A and C.

Now turning to the VB states, the conductance map in Fig. 3(e) was taken along a line spanning a location sequence A-B-C-A [as indicated in Fig. 3(d)] at energies around the VB band edge. In this case, confined states are observed at locations B and C – along with the aforementioned TIBB-induced streaks – and again no confinement occurs at location A. On the other hand, the minimum in VB band-edge position at location A (cf. Zhang et al.[7] and Table 1) is clearly visible. Note that the conductance associated with the VB fades out at locations B and C because of the height modulation in the moiré pattern [Fig. 1(d) and Table 1] and the fact that the mapping in Fig. 3(e) was performed at constant tip height. The spectra in Fig. 3(f) complement the measurement in the VB-band edge region and are in agreement with the conductance map in Fig. 3(e).

Before turning to an explanation for our observed band-edge spectral peaks, we first consider the fact that such features were not reported by Zhang et al.[7] Their studies were conducted at liquid nitrogen temperature (near 77 K), and so to investigate the possible influence of temperature we also performed studies at 80 K. The results are pictured in Fig. 4. We again observe the moiré pattern over the surface, Fig. 4(a), and STS measurements at various locations throughout the moiré unit cell reveal shifts in the energies of band edges, Fig. 4(b), i.e. similar to those seen both by Zhang et al. and in our 5 K measurements. Expanded views of the band edges are shown in the insets of Fig. 4(b) (also including a few additional spectra from nearby locations). Examining the edge of the $\Gamma_W$ VB (left inset), we find no trace of the sharp spectral features that were found in the 5 K data [Fig. 2(b), left inset]. For the $K_M$ CB, some "sharpening" of the band edge (a slight peak) is still apparent in the 80 K spectra [Fig. 4(b), right inset], but we do not consider this modest effect to be inconsistent with the prior work.[7] Comparison with the 5 K spectra of Fig. 2(b) reveals that significant broadening of the band-edge features has also occurred for these $K_M$ band-edge states. It should be noted that the 80 K temperature corresponds to $kT = 6.9$ meV, yielding an energy resolution[33] of $\Delta E = \sqrt{(3.5kT)^2 + (2.5eV_{mod})^2} = 35$ meV, which is only moderately larger than the 25 meV resolution for the 5 K data. Hence, the additional broadening apparent in the 80 K data appears to be considerably larger than expected from either instrumental effects or thermal occupation of electronic states. As a possible mechanism for this broadening, we note that low-energy phonon modes associated with TMD multilayers[34] might, through deformation-potential coupling, act to inhibit the formation of the sharp band-edge states; additional temperature-dependent work is needed to further elucidate this effect. In any case, consistent with these changes in the STS, Fig. 4(d) shows constant-height conductance mapping of the CB edge at 80 K (additional data is provided in Fig. S7). Spatial confinement of the states in the moiré unit cell is clearly evident, but in contrast to the 5 K results of Fig. 2(d), we now find no distinct features that we can associate with discrete quantum states.

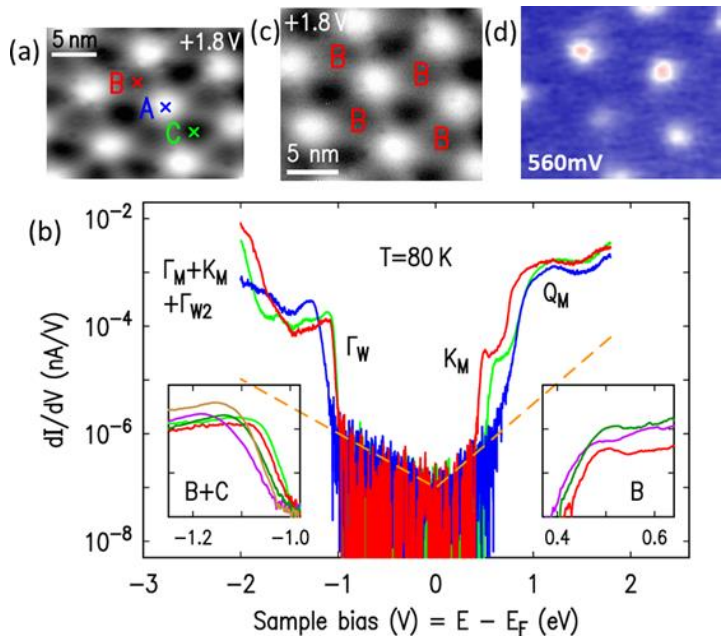

FIG 4. STM/STS data acquired at 80 K, showing (a) constant-current image of hetero-bilayer of $MoS_2$ on $WSe_2$, (b) tunneling spectra acquired at the locations indicated in (a), with insets showing expanded view of band-edge peaks and including data from nearby spatial locations, (c) constant-current image of hetero-bilayer, and (d) constant-height conductance map acquired at 560 mV (same color range as in Fig. 3(b)).

Let us now consider the origin of the sharp band-edge spectral peaks seen in the 5 K data. We apply similar reasoning as used for describing STS spectra arising from the moiré pattern of graphene on hexagonal boron nitride (h-BN),[35,36] although with one very important difference being that, for the $MoS_2$-$WSe_2$ system, we are dealing with bands with parabolic dispersion. We construct an effective potential for each band edge in the *full* moiré unit cell, using the band-edge energies of the small, *1×1-unit-cell* computations of Table 1 to estimate the effective potential. This spatially varying potential is constructed as a Fourier series, using *only* the Fourier components at the lowest nonzero reciprocal lattice vectors (potential term $V_\mathbf{G}$ on each of three equivalent **G** vectors, and $V_\mathbf{G}^*$ on the inequivalent set of three **G** vectors).[36] With this effective potential, we then solve the Schrodinger equation for the full moiré unit cell; with only six nonzero $V_\mathbf{G}$ terms, the problem is identical to that of a nearly free electron (NFE) model on a 2D hexagonal lattice, except that the "perturbing" $V_\mathbf{G}$ terms of the NFE model are now relatively large (compared to the "unperturbed" band widths) for the moiré problem. Assuming an effective mass of unity, then for our 8.5-nm moiré period, the dispersion of the lowest band is only 9.3 meV (e.g. out to the edge of the BZ at the K-point), neglecting the spatially varying terms in the potential. Then, including these terms, the well-known gaps of $2|V_\mathbf{G}|$ form at the BZ edges, yielding a band width for the lowest band that is *less* than this 9.3-meV value. As $|V_\mathbf{G}|$ increases, the width of the lowest band decreases and it shifts to lower energies, as shown in Fig. 5.

Concerning the size of the $|V_\mathbf{G}|$ terms,[36] an important contribution, already discussed by Zhang et al. in connection with their observed band shifts,[7] arises from hybridization of the states between $MoS_2$ and $WSe_2$. In particular, the $\Gamma_W$ state derived from the $WSe_2$ Γ-point VB is significantly perturbed by the adjoining $MoS_2$, producing a higher lying state (i.e. nearer to the VB edge) at the corrugation minimum compared to a maximum. Table 1 lists the corresponding energies of this band edge relative to the vacuum level on the $WSe_2$ side of the bilayer (the EG below the hetero-bilayer is known to have large n-type doping,[37] so using the electrostatic potential energy below the $WSe_2$ as a reference is appropriate). Considering the variation of this band-edge energy over the moiré cell, with the method of Jung et al.[36] we obtain $|V_\mathbf{G}^{\Gamma_W}| = 21$ meV; the NFE band structure of this case, as shown in Fig. 5, has a width of the lowest lying band of only 0.4 meV. We associate this band with the sharp (resolution limited) band-edge VB states seen in Figs. 3 and 4. Figure 5 (right) shows the wave function at the $\Gamma$-point associated with this lowest band. We find strong confinement of the state within the moiré unit cell, consistent with experiment.

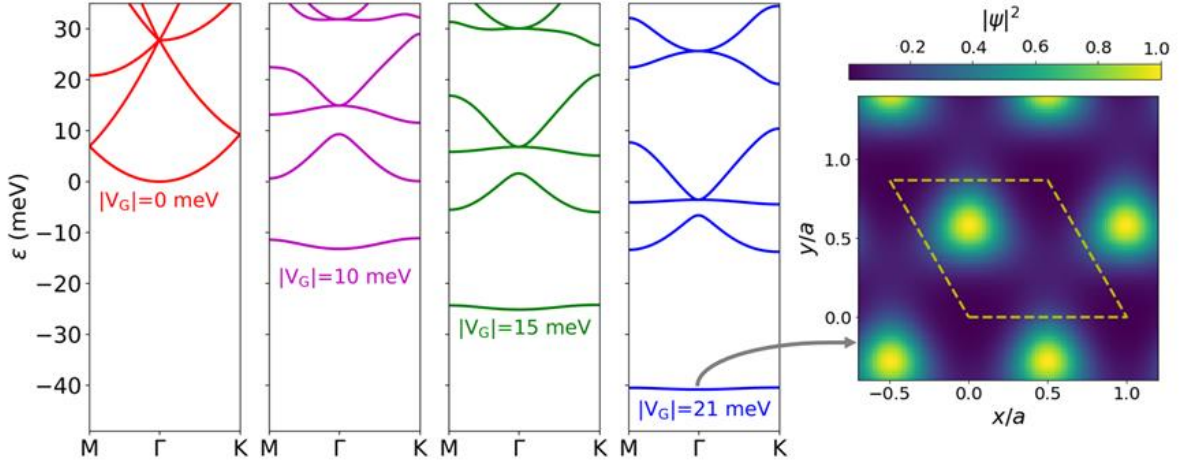

FIG 5. Left: Band structures for an NFE model on a hexagonal moiré lattice, employing different magnitudes for the $V_\mathbf{G}$ potential terms of the six lowest (nonzero) $\mathbf{G}$ vectors, $|V_\mathbf{G}|$. Right: Wave function for the state of the lowest band, at the $\Gamma$ point, for $|V_\mathbf{G}| = 21$ meV.

The confined states shown in Fig. 5 provide a qualitative explanation for the spectral peaks that we observe near the $\Gamma_W$ VB edge, except that in experiment these states are observed at both of the B and C corrugation minima [Fig. 3(b)] whereas in our theory described thus far we expect such states only at the C minima (Table 1, third column, with the VB $\Gamma_W$ band-edge energy at the C minimum being highest). However, we find nearly the same energies (within 50 meV) for the B and C minima, and we note that the prior theoretical results of Zhang et al.[7] (which include strain corrections) find the energies at the two minima to be equal within 20 meV with the B minima energy being highest.

Turning to the origin of the spectral peaks observed near the $K_M$ CB edge, the energies of the effective potential for this band edge are listed in the final column of Table 1. We see that the potential has nearly equal (within 20 meV) minima at the A and B locations, contrary to experiment in which the states are clearly confined to the B locations. However, we feel that detailed electrostatic modeling of the $MoS_2$-$WSe_2$-EG system, including the presence of the STM probe-tip, is necessary to fully understand the potential variation across the moiré unit cell. In particular, intrinsic (polarization) dipoles exist across the $MoS_2$-$WSe_2$ bilayer, with substantial variation in dipole energy across the unit cell (0.076, 0.165, 0.071 eV at the A, B, and C locations, respectively, where in all cases the electronic energy *increases* from the $WSe_2$ to the $MoS_2$ layer). Bound charge associated with the dipoles will be screened (nonlinearly) by induced free charge in the EG, which will produce further variations in energy across the unit cell. This screening effect of the EG is affected by its electron doping,[37] which will be large for sample bias $\leq 0$ but relatively small for bias $\gg 0$. Such effects must be modeled in detail before the locations of the quantum-confined states can be fully understood.

In summary, at a temperature of 5 K we observe single-particle quantum-confined states associated with the $\Gamma_W$ VB and $K_M$ CB in a $MoS_2$-$WSe_2$ hetero-bilayer. Such states have not been previously experimentally reported in TMD hetero-bilayers to our knowledge (although their presence is implicit in an early theoretical study,[9] as well as in two recent theoretical investigations of multi-particle effects[38,39]), but we believe that they are a general property of such systems. The states in our experiments turn out to be confined at the minima of the moiré corrugation, but this need not be the case for other bilayers (or even for the $MoS_2$-$WSe_2$ in the absence of a biased STM probe-tip and/or an underlying EG layer). The method of analysis we propose is the same as previously employed for graphene on h-BN,[36] with one crucial difference being that it is parabolic bands in the $MoS_2$ and $WSe_2$ that are perturbed by the moiré (unlike the highly dispersive, linear bands of graphene), leading to the confined states in the $MoS_2$-$WSe_2$ hetero-bilayer. We observe resolution-limited spectral peaks for the $\Gamma_W$ band-edge states in particular, with significant variation in those energies between moiré unit cells, suggestive of spatial localization of the states within a single cell. We also observe a rather large temperature dependence for these states, leading to their absence in observations at 80 K. As already suggested by Zhang et al., the large modulation in band edge positions observed for these TMD hetero-bilayers may be relevant for device application.[7] For example, the sharp, localized band edge states observed here (at 5 K), in analogy with localized states of coupled quantum dots, might serve as a useful platform for quantum computation.[40] On the other hand, the band-edge shifts observed both at 5 K and higher temperatures may be detrimental for application of such hetero-bilayer in interlayer tunneling devices.[41] Additional work is required to more fully determine the impact of the band-edge shifts and concomitant band-edge states in device applications.

We gratefully acknowledge discussions with D. Xiao, M. Widom, and V. Bheemarasetty (all of CMU), M. Hybertsen (Brookhaven), and V. Meunier (Rennselaer). This work was supported in part by the A. von Humboldt Foundation and by the Center for Low Energy Systems Technology (LEAST), one of six centers of STARnet, a Semiconductor Research Corporation program sponsored by Microelectronics Advanced Research Corporation (MARCO) and Defense Advanced Research Projects Agency (DARPA).

[1] Geim, A. K.; Grigorieva, I. V. *Nature* **2013**, 449, 419-425.
[2] Mak, K. F.; Lee, C.; Hone, J.; Shan, J.; Heinz, T. F. *Phys. Rev. Lett.* **2010**, 105, 136805.
[3] Wang, Q. H.; Kalantar-Zadeh, K.; Kis, A.; Coleman, J. N.; Strano, M. S. *Nat. Nanotechnol.* **2012**, 7, 699-712.
[4] Li, M.-Y.; Shi, Y.; Cheng, C.C.; Lu, L.-S.; Lin, Y.-C.; Tang, H.-L.; Tsai, M.-L.; Chu, C.-W.; Wei, K.-H.; He, J.-H.; Chang, W.-H.; Suenaga, K.; Li, L.-J. *Science* **2015**, 349, 524-528.
[5] Yun, W. S.; Han, S. W.; Hong, S. C.; Kim, I. G.; Lee J. D. *Phys. Rev. B* **2012**, 85, 033305.
[6] Chiu, M.-H.; Li, M.-Y.; Zhang, W.; Hsu, W.-T.; Chang, W.-H.; Terrones, M.; Terrones, H.; Li, L.-J. *ACS Nano* **2014**, 8, 9649-9656.
[7] Zhang, C.; Chuu, C.-P.; Ren, X.; Li, M.-Y.; Li, L.-J.; Jin, C.; Chou, M.-Y.; Shih, C.-K.; *Sci. Adv.* **2017**, 3, e1601459.
[8] Lin, Y.C.; Ghosh, R. K.; Addou, R.; Lu, N.; Eichfeld, S. M.; Zhu, H.; Li, M.-Y.; Peng, X.; Kim, M. J.; Li, L.-J.; Wallace, R. M.; Datta S.; Robinson, J. A. *Nat. Comm.* **2015**, 6, 7311.
[9] Kang, J.; Li, J.; Li, S.-S.; Xia, J.-B.; Wang, L.-W. *Nano Lett.* **2013**, 13, 5485-5490.

[10] Tongay, S.; Fan, W.; Kang, J.; Park, J.; Koldemir, U.; Suh, J.; Narang, D. S.; Liu, K.; Ji, J.; Li, J.; Sinclair, R.; Wu, J.; *Nano Lett.* **2014**, 14, 3185-3190.

[11] Kang, J.; Tongay, S.; Zhou, J.; Li, J.; Wu J. *Appl. Phys. Lett.* **2013**, 102, 012111.

[12] Chiu, M.-H.; Zhang, C.; Shiu, H.-W.; Chuu, C.-P.; Chen, C.-H.; Chang, C.-Y. S.; Chen, C.-H.; Chou, M.-Y.; Shih, C.-K.; Li, L.-J.; *Nat. Commun.* **2015**, 6, 7666.

[13] Rivera, P.; Seyler, K. L.; Yu, H.; Schaibley, J. R.; Yan, J.; Mandrus, D. G.; Yao, W.; Xu, X. *Science* **2016**, 351, 688-691.

[14] Rivera, P.; Schaibly, J. R.; Jones, A. M.; Ross, J. S.; Wu, S.; Aivazian, G.; Klement, P.; Seyler, K.; Clark, G.; Ghimire, N. J.; Yan, J.; Mandrus, D. G.; Yao, W.; Xu, X. *Nature Comm.* **2015**, 6, 6242.

[15] Hong, X.; Kim, J.; Shi, S.-F.; Zhang, Y.; Jin, C.; Sun, Y.; Tongay, S.; Wu, J.; Zhang, Y.; Wang, F. *Nat. Nanotechnol.* **2014**, 9, 682-686.

[16] Hermann, K. *J. Phys.: Condens. Matter* **2012**, 24, 314210.

[17] Hunt, B.; Sanchez-Yamagishi, J. D.; Young, A. F.; Yankowitz, M.; LeRoy, B. J.; Watanabe, K.; Taniguchi, T.; Moon, P.; Koshino, M.; Jarillo-Herrero, P.; Ashoori, R. C. *Science* **2013**, 340, 1427-1430.

[18] Kumar, H.; Er, D.; Dong, L.; Li, J.; Shenoy, V. B. *Sci. Rep.* **2015**, 5, 10872.

[19] Dean, C. R.; Wang, L.; Maher, P.; Forsythe, C.; Ghahari, F.; Gao, Y.; Katoch, J.; Ishigami, M.; Moon, P.; Koshino, M.; Taniguchi, T.; Watanabe, K.; Shepard, K. L.; Hone, J.; Kim, P. *Nature* **2013**, 497, 598-602.

[20] Xue, J.; Sanchez-Yamagishi, J.; Bulmash, D.; Jacquod, P.; Deshpande, A.; Watanabe, K.; Taniguchi, T.; Jarillo-Herrero, P.; LeRoy, B. J. *Nat. Mater.* 2011, 10, 282-285.

[21] Kim, H. S.; Gye, G.; Lee, S.-H.; Wang, L.; Cheong, S.-W.; Yeom, H. W. *Sci. Rep.* 2017, 7, 12735.

[22] Wu, F.; Lovorn, T.; MacDonald, A. H.; *Phys. Rev. Lett.* **2017**, 118, 147401.

[23] Eichfeld, S. M.; Colon, V. O.; Nie, Y.; Cho, K.; Robinson, J. A.; *2D Mater.* **2016**, 3, 25015.

[24] Subramanian, S.; Deng, D. D.; Xu, K.; Simonson, N.; Wang, K.; Zhang, K. H.; Li, J.; Feenstra, R.; Fullerton-Shirey, S. K.; Robinson, J. A. *Carbon* **2017**, 125, 551.

[25] Lin, Y.-C.; Jariwala, B.; Bersch, B. M.; Xu, K.; Nie, Y.; Wang, B.; Eichfeld, S. M.; Zhang, X.; Choudhury, T. H.; Pan, Y.; Addou, R.; Smyth, C. M.; Li, J.; Zhang, K.; Haque, M. A.; Fölsch, S.; Feenstra, R. M.; Wallace, R. M.; Cho, K.; Fullerton-Shirey, S. K.; Redwing, J. M.; Robinson, J. A. *ACS Nano* **2018**, to appear, DOI: 10.1021/acsnano.7b07059

[26] Mårtensson, P.; Feenstra, R. M. *Phys. Rev. B* **1989**, 39, 7744.

[27] Kresse, G.; Furthmüller, J. *Phys. Rev. B* **1996**, 54, 11169.

[28] Kresse, G.; Joubert, D. *Phys. Rev. B* **1999**, 59, 1758.

[29] Perdew, J. P.; Burke, K.; Ernzerhof, M. *Phys. Rev. Lett.* **1996**, 77, 3865; erratum, *Phys. Rev. Lett.* **1997**, 78, 1396.

[30] Grimme, S. *J. Comp. Chem.* **2006**, 27, 1787.

[31] Stroscio, J. A.; Feenstra, R. M.; Fein, A. P. *Phys. Rev. Lett.* **1986**, 57, 2579.

[32] Hill, H. M.; Rigosi, A. F.; Rim, K. T.; Flynn, G. W.; Heinz, T. F. *Nano Lett.* **2016**, 16, 4831-4837.

[33] Morgenstern, M. *Surf. Rev. Lett.* **2003**, 10, 933.

[34] Zhao, Y.; Luo, X.; Li, H.; Zhang, J.; Araujo, P. T.; Gan, C. K.; Wu, J.; Zhang, H.; Quek, S. Y.; Dresselhaus, M. S.; Xiong, Q. *Nano Lett.* **2013**, 13, 1007−1015.

[35] Yankowitz, M.; Xue, J.; Cormode, D.; Sanchez-Yamagishi, J. D.; Watanabe, K.; Taniguchi, T.; Jarillo-Herrero, P.; Jacquod, P.; LeRoy, B. J. *Nat. Phys.* **2012**, 8, 382-386.

[36] Jung, J.; DaSilva, A. M.; MacDonald, A. H.; Adam, S. *Nat. Comm.* **2015**, 6, 6308.

[37] Ristein, J.; Mammadov, S.; Seyller, Th. *Phys. Rev. Lett.* **2012**, 108, 246104.

[38] Yu, H.; Liu, G.-B.; Tang, J.; Xu, X.; Yao, W.; *Sci. Adv.* **2017**, 3, e1701696.

[39] Wu, F.; Lovorn, T.; MacDonald, A. H. *Phys. Rev. Lett.* **2017**, 118, 147401.

[40] Eriksson, M.; Coppersmith, S. N.; Lagally, M. G. *MRS Bulletin* **2013**, 38, 794-801.

[41] Li, J.; Nie, Y.; Cho, K.; Feenstra, R. M. *J. Electron. Mater.* **2016**, 46, 1378.

## Supporting Information:

## Sample preparation

The MOCVD growth of WSe$_2$ is performed at 700 Torr using H$_2$ as a carrier gas at 800 °C, with W(CO)$_6$ and H$_2$Se precursors being introduced separately into a cold wall vertical reactor chamber and their respective flow rates controlled via mass flow controllers. The optimized condition for the growth was based on a recent detailed study of WSe$_2$ growth.[1] On top of these layers MoS$_2$ is deposited by CVD, using 2 mg MoO$_3$ and 200 mg sulfur powder as the optimal precursor ratio for synthesis performed at 850 °C. The substrate consists of epitaxial graphene (EG) formed on SiC.

## Scanning tunneling microscopy/spectroscopy (STM/STS)

The STM/STS measurements were carried out with a cryogenic STM operated in ultrahigh vacuum at 5 K or 80 K, as indicated in the text. Electrochemically etched tungsten tips cleaned in UHV by Ne ion bombardment and electron beam heating were used. STM images were recorded in constant-current mode using currents in the range 0.01 – 0.1 nA; bias voltages refer to the sample with respect to the STM tip. STS measurements of the differential tunneling conductance *dI/dV* were carried out with lock-in technique (modulation frequency 675 Hz at a peak-to-peak modulation of 10 mV unless otherwise specified) to probe the local density of electronic states. We employ a variable-*z* measurement method in which an offset Δ*S*(*V*), which varies linearly with the magnitude of the sample bias *V*, is applied to the tip-sample separation.[2] The exponential increase in conductance arising from this variation in tip-sample separation is then normalized by multiplying the data by a factor of $e^{2\kappa \Delta S(V)}$, where $\kappa$ is an experimentally determined decay constant of 10 nm$^{-1}$ (averaged over bias voltage). This measurement method and subsequent normalization does not affect any detailed structure in the spectra, but it improves the dynamic range by 1 – 2 orders of magnitude. The noise level for the conductance is also measured, and normalization of that using the same method then yields a voltage-dependent noise level for each spectrum. Band edges are determined simply by the voltage (energy) at which the observed band-edge conductance intersects the noise level (or the observed conductance of the underlying graphene layer); see Fig. S4. This method is perhaps somewhat qualitative compared to the detailed fitting method of Hill et al.,[3] but on the other hand, that fitting method explicitly does *not* include the type of band edge shifts (and the associated somewhat *gradual* turn-on of the conductance at the band edges) observed both by Zhang et al.[4] and in the present work.

## Theoretical Modeling

The density functional theory calculations are performed as described in the main body of the manuscript. 1x1 unit cells of $MoS_2$ on $WSe_2$ with various translational registrations between the $MoS_2$ and $WSe_2$ are computed, using lattice parameter of 3.25 Å and relaxation of all atoms. The wave functions are expanded in plane waves with a cutoff energy of 400 eV, and the energy convergence criteria for electronic and ionic optimization are $10^{-4}$ eV and 0.01 eV/Å, respectively. Integration over the first Brillouin zone is carried out with a Γ-centered 24×24×1 **k**-point mesh for the wave function calculations. Spin-orbit coupling is employed in the calculation. A vacuum region of over 10 Å in the direction normal to the 2D material layers is added to minimize the interaction between the adjacent supercell images.

## Lattice orientations of $MoS_2$ and $WSe_2$ layers

The lattice orientations of $MoS_2$ and $WSe_2$ lattices are analyzed on the basis of a large number of atomic resolution STM images. We find the majority of the sample to have rotational misalignment that is either 0° (R-stacking) or 180° (H-stacking). We are unable to distinguish between these two possibilities on the basis of STM imaging alone, and as commented in the main manuscript, we tentatively accept the identification of the 0° stacking made by Zhang et al. on the basis of annular dark-field scanning transmission electron microscopy.[4] A representative example of our stacking determination is shown in Fig. S1. The hexagonal moiré pattern arising from the $MoS_2$-$WSe_2$ vertical hetero-bilayer appears on a triangular $MoS_2$ island, as shown in Figs. S1(a) and (b). The lattices of $WSe_2$ and $MoS_2$ are shown in the atomic resolution images in Figs. S1(c) and (d). The $MoS_2$ and $WSe_2$ lattices are in the same orientation, as indicated by the dashed red line.

Most surface locations are found to display the same stacking arrangement as in Fig. S1, although defects in the moiré lattice are commonly observed (such as near the center of the triangular island of Fig. 1 of the main manuscript, where no moiré corrugation whatsoever is observed). In some cases, these sorts of defects in the moiré lattice are observed to be correlated with the occurrence of surface defects, including both point defects such as small apparent clusters (possibly surface contamination) and surface steps (most likely occurring in the underlying epitaxial graphene, with the $MoS_2$-$WSe_2$ bilayer appearing to extend uniformly, like a carpet, over the steps). We speculate that such surface defects then affect the subtle energetics of the moiré structure, with the 0° stacking being the low-energy equilibrium one that is formed during growth on defect-free surface areas,[5] but where surface defects can substantially perturb this moiré lattice. We expect that a moiré arrangement that differs from the one we generally observe will have a substantially different band-edge electronic structure, e.g., for 0° and 180° stacking arrangements we find quite different theoretical band-edge shifts (the former are listed in Table 1 of the main manuscript), but in any case our experimental studies to date have been limited to only well-ordered moiré areas, all of which appear to have the same type of stacking (i.e. 0°).

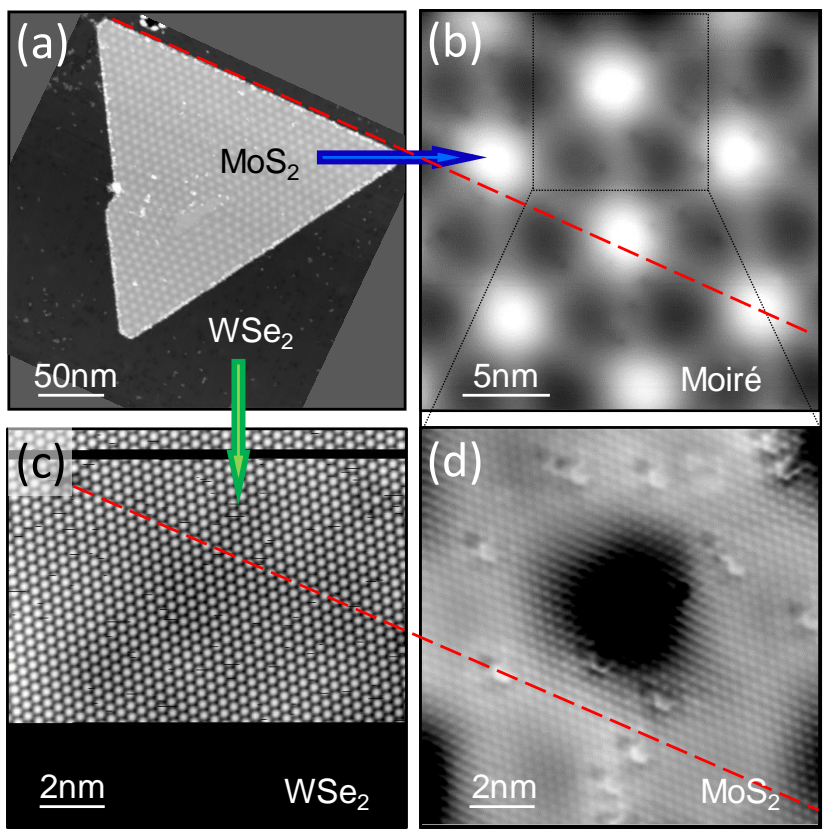

**Figure S1.** STM topography images of the $MoS_2$-$WSe_2$ vertical hetero-bilayer. (a) A large-scale image (2.5 V, 10 pA) showing a triangular monolayer $MoS_2$ island on top of monolayer $WSe_2$. It has been rotated 24° clockwise to compensate the -24° rotation during image acquisition. (b) Close-up image (2.0 V, 100 pA) taken on the $MoS_2$ island, showing the corrugated moiré. (c) Atomic resolution image (-1.0 V, 100 pA) of $WSe_2$. (d) Atomic resolution image (-1.1 V, 100 pA) of $MoS_2$, which is taken in the area in (b) marked by the rectangular. The dashed red lines are parallel to a close-packed row of atoms in the atomic resolution images, indicating the $MoS_2$ and $WSe_2$ lattices are in the same orientation.

# Bias-dependent images of the $MoS_2$-$WSe_2$ hetero-bilayer

In the bias range above the conduction band minimum (CBM, near 1.0 V) and below the valence band maximum (VBM, near -1.25 V) of location A, a regular moiré pattern is seen in bias-dependent images. When the bias is around the energy of the band gap edges, additional features in the location B are seen due to the confined electronic states. There are also some randomly distributed bright spots in the bias range from -0.4 V to -1.0 V, which we attribute to defect-induced gap states. In the bias range from 0.5 V to -0.4 V, a distinct pattern with alternating bright and dark triangles, each extending over half of the moiré unit cell, is seen. The origin of this contrast – which was also observed by Zhang et al.[4] – is not yet known.

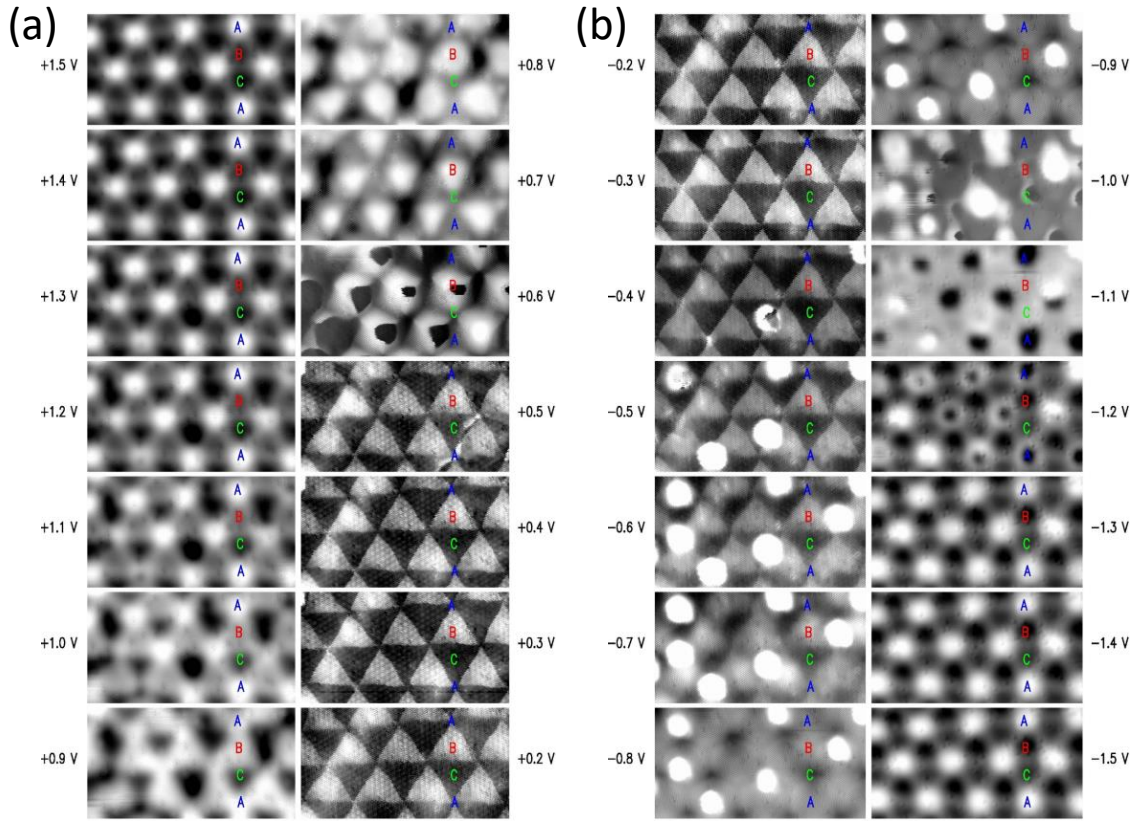

**Figure S2.** Bias-dependent constant-current STM images of the $MoS_2$-$WSe_2$ hetero-bilayer. (a) Positive bias series. (b) Negative bias series. Representative A, B and C locations in the moiré unit cell are indicated in each image.

## Structure model of the moiré pattern

The model of the moiré pattern is built by superimposing the $MoS_2$ lattice on top of the $WSe_2$ lattice. A unit cell of the moiré contains 27 unit cells of the $MoS_2$ and 26 of the $WSe_2$. Regarding the registry of the atoms, there are three special locations in a moiré unit cell, as indicated by A, B and C in Fig. S3, according to the registry of the atoms. At location A, the metal atoms (chalcogen) of the upper layer are over the metal (chalcogen) atoms of the lower layer. At location B, the Mo is over Se, and S and W are not vertically aligned with each other. Finally, at location C, the S is over W, whereas Mo and Se are not vertically aligned with each other. Our notation A, B and C corresponds to the notation AA, $AB_W$ and $AB_{Se}$ in the paper of Zhang et al.[4]

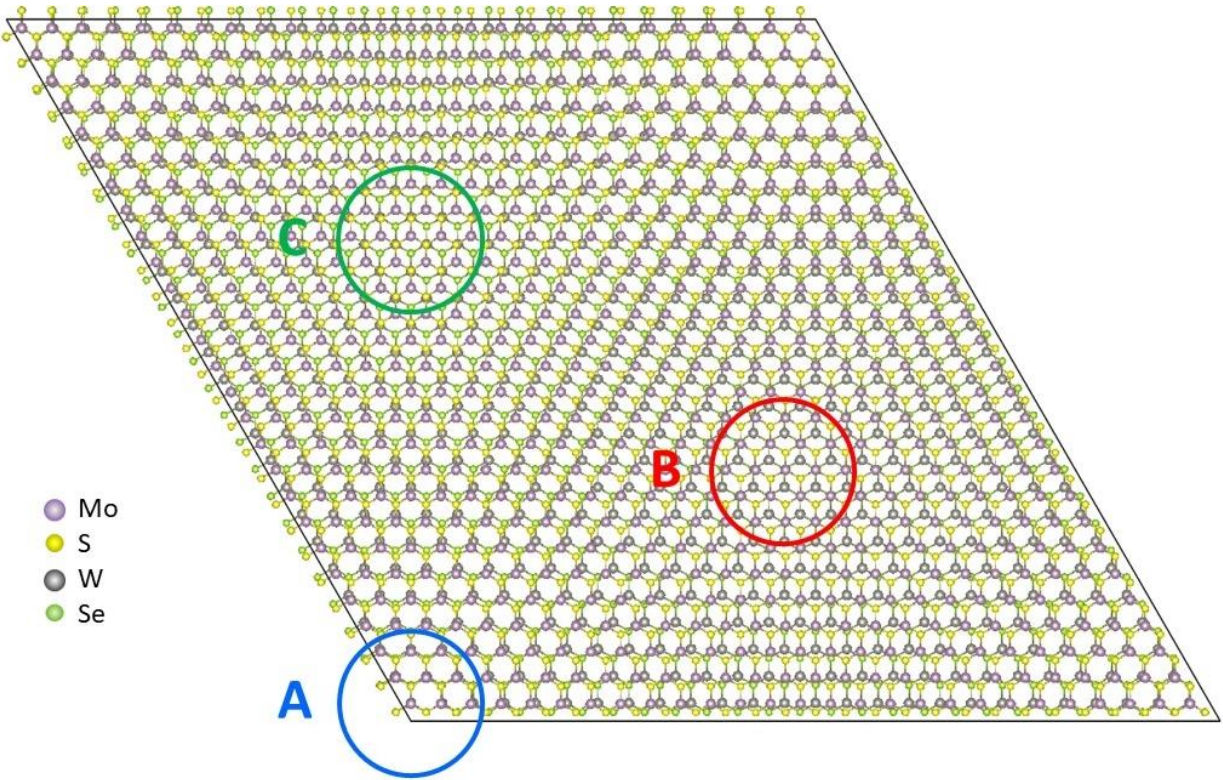

**Figure S3.** Ball-and-stick structure model of the $MoS_2$-$WSe_2$ hetero-bilayer. $MoS_2$ is shown on top, with Mo as purple balls and S as small yellow balls. $WSe_2$ is shown below, with W as gray balls and Se as small green balls. The three high-symmetry locations in the moiré unit cell, A, B and C, are indicated by large blue, red, and green circles, respectively.

## STS of isolated MoS$_2$ and WSe$_2$ layers, compared to hetero-bilayer

STS of individual MoS$_2$ and WSe$_2$ layers, acquired on corresponding samples, are shown in Figure S4 along with a spectrum obtained from the MoS$_2$-WSe$_2$ hetero-bilayer. Notation for the labeling of bands is same as in the main text. Band edges are indicated by the red lines, yielding measured gaps of 2.02±0.02 and 1.93±0.02 eV for MoS$_2$ and WSe$_2$, respectively, and 1.13±0.02 eV for the hetero-bilayer. Lock-in amplifier modulation voltage of $V_{rms} = 50$ mV was used in these measurements, for improved signal to noise. The modulation produces an upwards (downwards) shift of the valence (conduction) band edge, by an amount equal to the peak amplitude of the modulation, $\sqrt{2}\,V_{rms}$. Hence, the measured gaps must be increased by $2\sqrt{2}\,V_{rms}$, yielding corrected gaps of 2.16±0.02 and 2.07±0.02 eV for MoS$_2$ and WSe$_2$, respectively, and 1.27±0.02 eV for the hetero-bilayer. The spectra from the hetero-bilayer was acquired at a B-type corrugation minimum, thus revealing a spectral peak at the $K_M$ conduction band edge. No peak is seen at the $\Gamma_W$ valence band edge since the 50 mV modulation broadens this peak sufficiently so that it is unobservable. Note the clear observation of the $K_W$ valence band of WSe$_2$, seen both in the spectrum of the isolated WSe$_2$ layer and for the hetero-bilayer. Additionally, for the isolated WSe$_2$ layer, measurable conductance is seen across most of the band gap region. This conductance arises from the graphene layer underlying the WSe$_2$. It is routinely observed for monolayers of TMD materials on epitaxial graphene, so long as the measurements are performed with sufficient dynamic range in the conductance.

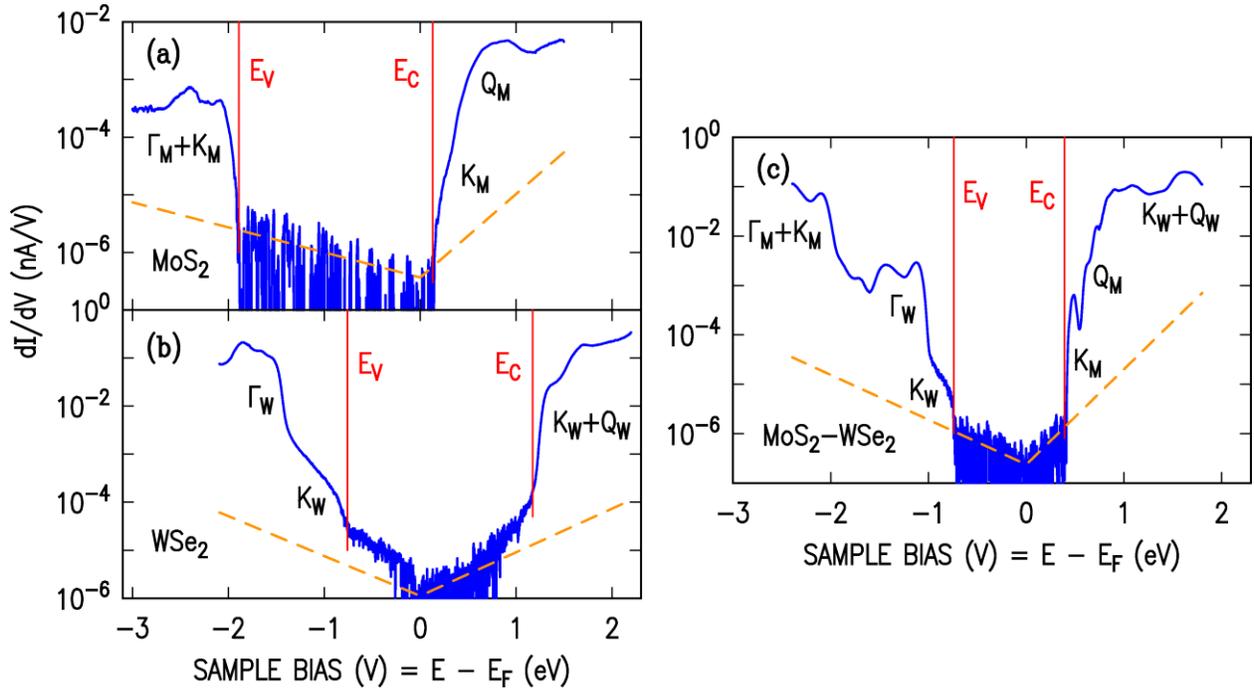

**Figure S4.** STS acquired at 5 K showing: Left: spectra obtained from a monolayer of MoS$_2$ (upper panel) and of WSe$_2$ (lower panel); Right: a spectrum obtained near a B-type corrugation minimum of the MoS$_2$-WSe$_2$ hetero-bilayer. All layers deposited on epitaxial graphene.

## Detailed STS of the MoS$_2$-WSe$_2$ hetero-bilayer

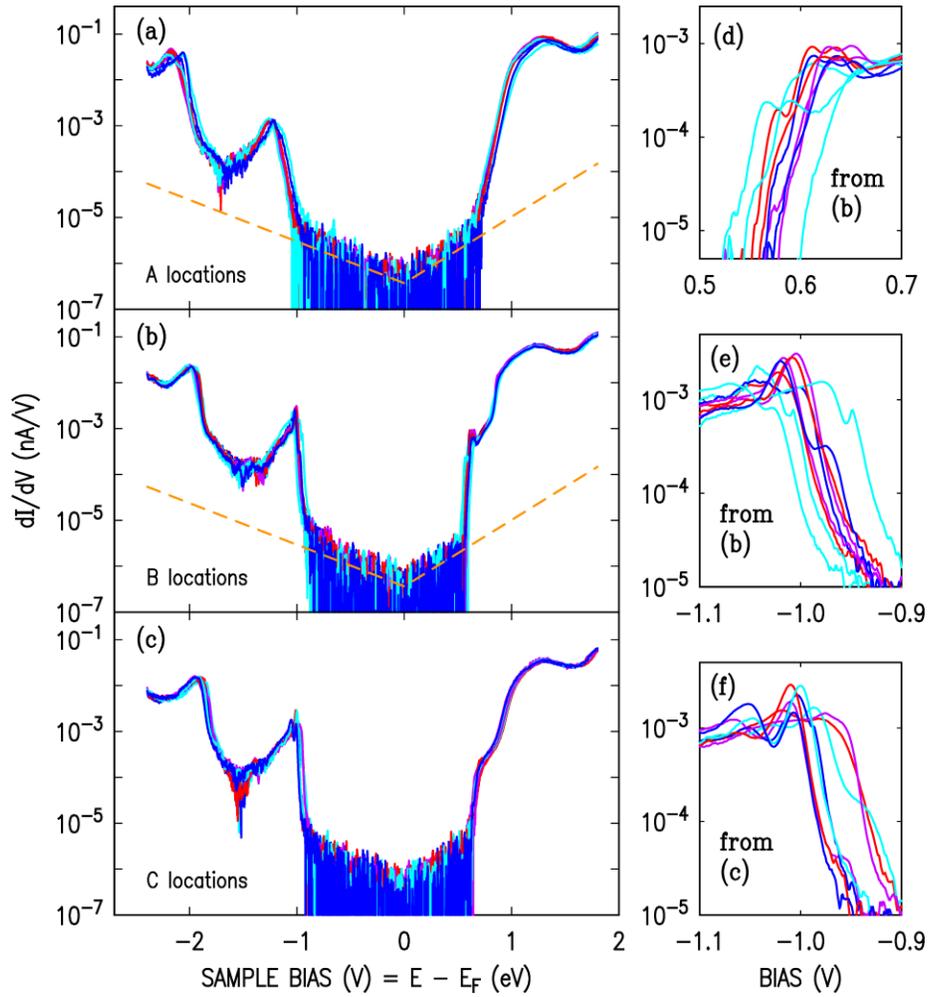

**Figure S5.** Detailed scanning tunneling spectra from a hetero-bilayer of MoS$_2$ on WSe$_2$, acquired at 5 K at locations indicated in the images in each panel, location types A – C for panels (a) – (c), respectively. (d) – (f) Expanded view of band edges.

# Constant-height conductance image series at 5 K

The spatial distribution and evolution of the confined electronic states around the CB and VB edges are measured by taking constant-height conductance maps at varying bias voltages. For the CB edge, B-centered rings of enhanced conductance emerge at ~500 mV. With increasing bias, these rings shrink in diameter and eventually merge into pronounced maxima. Ring shape and onset energy vary slightly from one to another B location, which we attribute to a perturbation arising from randomly distributed point defects. The perturbation implies a slight detuning in the confining potential at adjacent B locations, even though the $MoS_2$-$WSe_2$ hetero-bilayer is laterally continuous. Each B location is equivalent to a quasi quantum dot (QD). Consequently, a stable hexagonal array of quasi QDs has formed on the $MoS_2$-$WSe_2$ hetero-bilayer in a certain energy range (500 to 700 mV in this case). In the energy range around the VBM, confined states exist at both B and C locations.

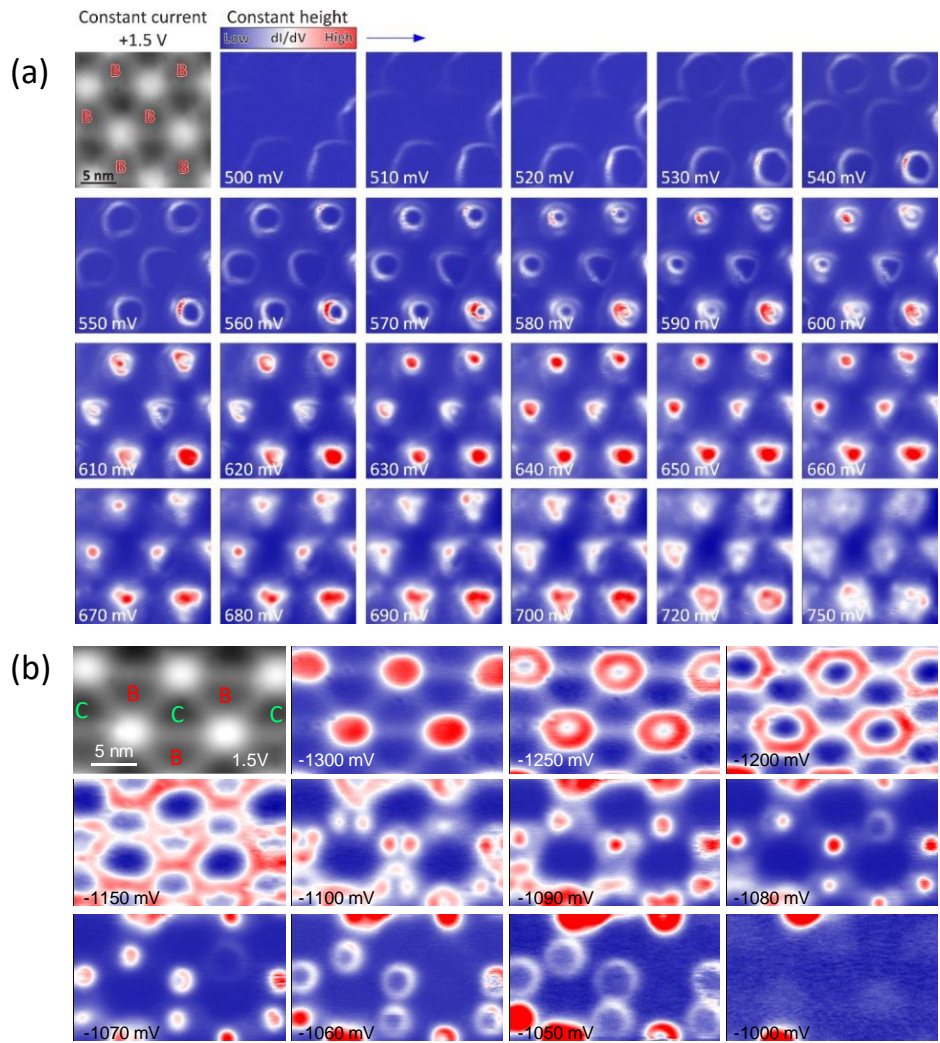

**Figure S6.** Gray-scale panels: topography images; Color-coded panels: constant-height conductance maps taken at the same area at varying biases. (a) 500 to 750mV; (b) -1300 to -1000mV. The data are acquired at 5 K.

## Constant-height conductance image series at 80 K

Constant-height conductance maps taken at 80 K indicate that the confined states still exist at such temperature. However, the fine structure in the conductance maps is lost due to the reduced energy resolution (Fermi broadening) in STS measurements at higher temperature.

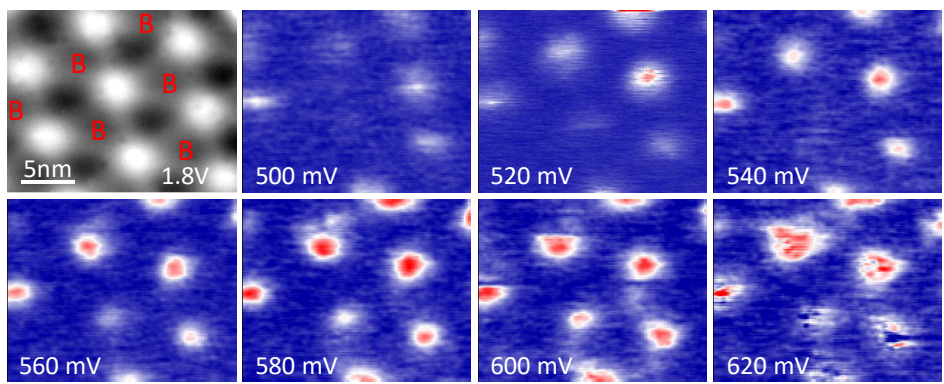

**Figure S7.** Upper left panel, STM image (21 nm × 16.7nm) of the $MoS_2$-$WSe_2$ moiré structure with the locations B indicated. All other panels: constant-height conductance maps taken at the same area at bias voltages from 500 to 620 mV. The data are acquired at 80 K.

## References

[1] Y.-C. Lin, B. Jariwala, B. M. Bersch, K. Xu, Y. Nie, B. Wang, S. M. Eichfeld, X. Zhang, T. H. Choudhury, Y. Pan, R. Addou, C. M. Smyth, J. Li, K. Zhang, M. A. Haque, S. Fölsch, R. M. Feenstra, R. M. Wallace, K. Cho, S. K. Fullerton-Shirey, J. M. Redwing, J. A. Robinson, ACS Nano, to appear (2018), DOI: 10.1021/acsnano.7b07059

[2] P. Mårtensson and R. M. Feenstra, Phys. Rev. B **39**, 7744 (1989).

[3] H. M. Hill, F. Rigosi, K. T. Rim, G. W. Flynn, and T. F. Heinz, Nano Lett. **16**, 4831 (2016).

[4] C. Zhang, C.-P. Chuu, X. Ren, M.-Y. Li, L.-J. Li, C. Jin, M.-Y. Chou, C.-K. Shih, Sci. Adv. **3**, e1601459 (2017).

[5] G. C. Constantinescu and N. D. M. Hine, Phys. Rev. B **91**, 195416 (2015).